\documentclass[twocolumn,prl,showpacs]{revtex4-1}
\usepackage{epsf,graphicx,amssymb}

\usepackage{epstopdf}
\usepackage{amsmath,amssymb}  
\usepackage{bbold}
\usepackage{mathbbol}
\usepackage{amscd}
\usepackage{amsfonts} 
\usepackage{graphicx} 
\usepackage{color}

\usepackage{amssymb,amsmath,amsthm}
\usepackage[colorlinks=true,linkcolor=blue,citecolor=blue]{hyperref}

\begin{document}
\title{Slanted snaking of localized Faraday waves}
\author{Basti\'an Pradenas}
\author{Isidora Araya}
\author{Marcel G. Clerc}
\author{Claudio Falc\'on}
\email{cfalcon@ing.uchile.cl}
\address{Departamento de F\'isica, Facultad de Ciencias F\'isicas y Matem\'aticas, Universidad de Chile, Casilla 487-3, Santiago, Chile}
\author{Punit Gandhi}
\author{Edgar Knobloch}
\email{knobloch@berkeley.edu}
\address{Department of Physics, University of California, Berkeley, California 94720, USA}
\date{\today}


\begin{abstract}

We report on an experimental, theoretical and numerical study of slanted snaking of spatially localized parametrically excited waves on the surface of a water-surfactant mixture in a Hele-Shaw cell. We demonstrate experimentally the presence of a hysteretic transition to spatially extended parametrically excited surface waves when the acceleration amplitude is varied, as well as the presence of spatially localized waves exhibiting slanted snaking. The latter extend outside the hysteresis loop. We attribute this behavior to the presence of a conserved quantity, the liquid volume, and introduce a universal model based on symmetry arguments, which couples the wave amplitude with such a conserved quantity. The model captures both the observed slanted snaking and the presence of localized waves outside the hysteresis loop, as demonstrated by numerical integration of the model equations.
\end{abstract}

\date{\today}
\maketitle

Dissipative systems driven out of equilibrium display spatial patterns that can extend over the whole system or be confined to a localized region~\cite{Descalzi2011,Purwins2010,CrossHohenberg1993}. Examples of the latter include dissipative solitons, fronts and localized oscillations called oscillons~\cite{CrossHohenberg1993} and these have been observed in a great variety of physical systems ranging from simple and complex fluids to liquid crystals, chemical reactions, magnetic materials, granular media and elastic solids. The theoretical understanding of these structures is largely based on detailed studies of the bistable Swift--Hohenberg equation~\cite{Knobloch2015} although geometrical methods have been successfully applied to one-dimensional systems~\cite{Coullet2000,Beck}. This understanding centers on the presence of a pinning mechanism whereby fronts connecting a homogeneous state to a spatially structured state pin to the latter, resulting in the presence of stationary fronts even at parameter values away from the Maxwell point at which the structured state has the same energy as the competing homogeneous state~\cite{CrossHohenberg1993}. The theory shows that the resulting pinning region contains an infinite number of coexisting localized states organized in a snakes-and-ladders structure straddling the Maxwell point~\cite{Burke2006,Burke2007}. Such states are therefore located within the hysteresis loop between the homogeneous and structured states. For steady localized states the corresponding bifurcation diagrams can take the form of standard snaking \cite{Knobloch2015} in which different length states all coexist within a pinning region, or their intervals of existence may be staggered generating slanted snaking~\cite{Firth2007,Barbay2008,Abshagen2010,Haudin2011,Thiele2013,Lloyd2015}, an effect attributed to nonlocality~\cite{Firth2007b,Dawes2008}. Neither type of structure has thus far been reported in experiments on localized standing oscillations, even though such oscillations are known to be present within the hysteresis loop between extended standing waves and a homogeneous state~\cite{Assemat,Clerc,Alnahdi}.
\begin{figure}[t]
\centering
\includegraphics[width=1.0\columnwidth]{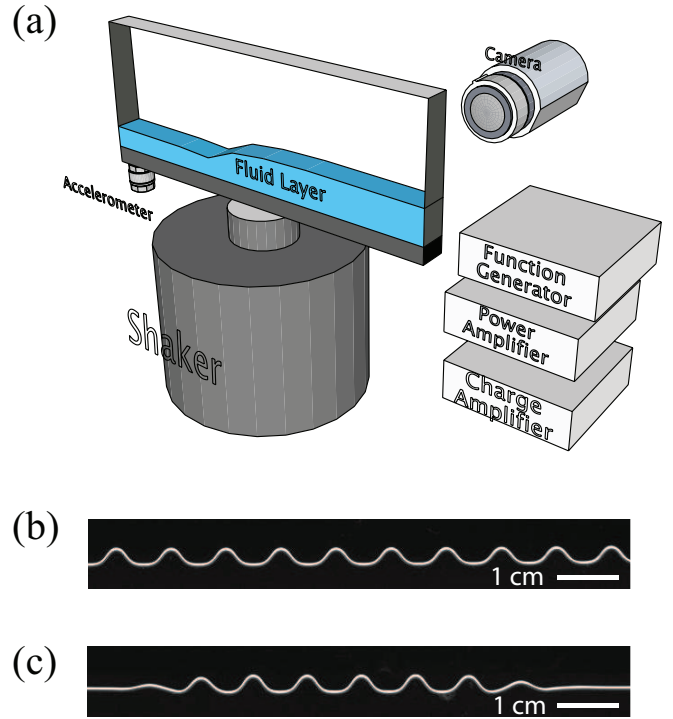} 
\caption{(color online) (a) Experimental setup showing a glass cell on a shaker. (b) Spatially extended Faraday wave. (c) Spatially localized Faraday wave with $n=8$ peaks. In (b) and (c) only a part of the whole domain is shown.}
\label{FigSetup}
\end{figure}

In this Letter we demonstrate slanted snaking of localized oscillations in a Faraday wave experiment~\cite{Faraday1831}, specifically waves at the surface of a water-surfactant mixture held between two-plates in a Hele-Shaw configuration and vibrated vertically. The extended wave pattern appears across the entire experimental cell through a subcritical instability which depends on the contact angle hysteresis of the triple line and the confinement width. Localized Faraday waves are observed both inside and outside of the hysteresis loop between the extended pattern and the flat interface, and display a snaking structure that is slanted. The organization of the observed localized oscillations can be understood using a pair of amplitude equations that capture the coupling between the amplitude of the parametrically excited standing waves and a long wave conserved mode representing the volume of liquid withheld by the meniscus. Numerical simulations of the proposed coupled equations are in good qualitative agreement with theoretical predictions and experimental observations.

\begin{figure}[b]
\centering
\includegraphics[width=1.0\columnwidth]{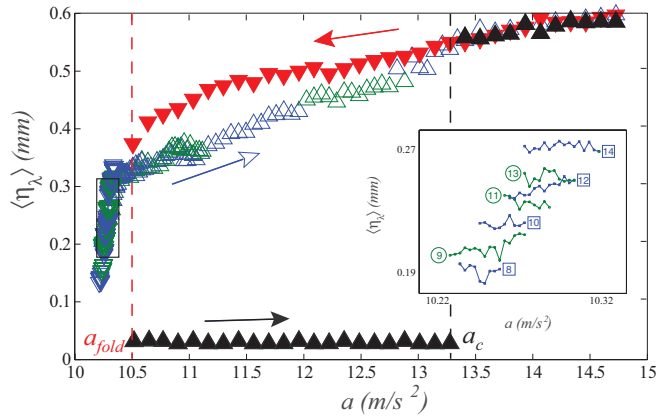} 
\caption{(color online) Experimental bifurcation diagram for a glass cell and increasing ($\blacktriangle$) and decreasing ($\blacktriangledown$) acceleration $a$, displaying a hysteresis loop in the pattern amplitude $\langle \eta_{\lambda}\rangle$ between $a_c$ and $a_{fold}$. Localized structures for increasing $(\bigtriangleup)$ and decreasing $(\bigtriangledown)$ $a$ are also displayed. The colors correspond  to even (blue) and odd (green) numbers of peaks $n$, $5\le n\le20$.  Inset: Detail of slanted snaking bifurcation diagram of localized structures for $n$ even (blue squares) and $n$ odd (green circles) outside the hysteresis loop for $8 \le n \le 14$. The results are suggestive of intertwined branches of odd and even localized states exhibiting slanted snaking.}\label{FigBif}
\end{figure}

The experimental setup is depicted in Fig.~\ref{FigSetup}. Two glass plates are separated by an aluminum frame creating a glass container (width $w=1$ mm, height $h=40$ mm, length $l=200$ mm) which is filled to a height $h_0=23$ mm with a mixture of 85$\%$ distilled water and 15$\%$ surfactant (Kodak Photo-Flo2 200\textsuperscript{TM}) by volume. The width $w$ and the height $h_0$ vary across the experimental cell by less than 5$\%$ and 1$\%$, respectively. The density $\rho$ of the working fluid is estimated at 1004 kg/m$^{3}$.  
Its surface tension $\sigma$ was measured at 22 mN/m by the sessile drop method~\cite{DelRioJColloid1997}. Its kinematic viscosity was measured at 3 $\pm$ 0.3 cSt  using a stress-controlled rheometer (RheolabQC from Anton-Paar) at room temperature ($20\pm2^{\circ}$C) in a Couette configuration (inner diameter 19.39 mm, outer diameter 21.00 mm, angle of the rotating cone 120$^{\circ}$) for steady shear rates  $\dot{\gamma}\in$[40,100] s$^{-1}$, as appropriate to our experiments. 
Normal stresses were not measured. The wetting angle of the fluid mixture on glass was measured between [5$^{\circ}$,15$^{\circ}$], showing partial wetting and contact angle hysteresis~\cite{DeGennesReview}. The glass container with the working fluid was mounted vertically on an electromechanical shaker driven sinusoidally by a function generator via a power amplifier. The vertical modulation of the acceleration $a(t)=a \cos{(\omega t)}$ was monitored by a piezoelectric accelerometer via a charge amplifier. Horizontal accelerations were less than 1$\%$. The experimental control parameter range was $a\in[2,20]$ m/s$^{2}$ and $f\equiv\omega/2\pi\in[30,60]$ Hz. For $f<35$ Hz, no  localized structures were found. In the present study we use $f=$ 45 Hz while $a$ spans the whole experimental range. 

Images of the surface profile are acquired with a CCD camera over a 10 s time window in a $1220\times200$ px spatial window (0.12 mm/px sensitivity in the horizontal direction and 0.11 mm/px in the vertical direction). For each value of $a$, a sequences of images is taken at frequency $f/2$ using the second output of the function generator as a trigger to ensure a stroboscopic view of the oscillating surface profile. 
The profile of the fluid elevation $\eta(x,t)\equiv h(x,t)-h_0$ is tracked for every point $x$ in space at each instant $t$ using a simple threshold intensity algorithm~\cite{MaciasPRE2013}, as shown in Fig.~\ref{FigSetup}. To do this, white light is passed through a diffusing screen from behind the layer while images are taken from the front, enhancing contrast and improving the functioning of the surface tracking algorithm. Figure~\ref{FigSetup} shows snapshots of the layer displaying (b) a spatially extended wave and (c) a spatially localized wave.

As we increase $a$ above the critical value $a_c$=13.35 m/s$^{2}$, the fluid layer becomes unstable to a parametric instability, displaying a stationary pattern of waves filling the entire experimental cell. Figure~\ref{FigSetup}(b) shows a part of this state. The pattern oscillates at frequency $f/2$ with onset wavelength $\lambda=9$ mm. The shape of the pattern is not symmetric with respect to the mean height: cusps are sharp and troughs are smooth and flat (see Fig.~\ref{FigSetup}). When $a$ is decreased the extended pattern persists down to $a_{fold}$= 10.5 m/s$^{2}<a_c$ where it collapses to the homogeneous flat state, indicating the presence of a hysteresis loop. We have checked that the critical wavelength $\lambda$ does not change with changes in $a$. To characterize the pattern evolution we compute $\langle\eta_{\lambda}\rangle$, the square root of the time and space-averaged mean-square local surface deformation. Figure~\ref{FigBif} shows $\langle \eta_{\lambda}\rangle$ as a function of $a$. We have confirmed that the hysteresis loop is reproducible by repeating the experimental procedure 5 times with a waiting time of 30 s between measurements. Below $a_{fold}$ our experimental system displays localized patterns, characterized by the number $n$ of peaks exhibited by the pattern (Fig.~\ref{FigSetup}(c)), with $n\in[5,N]$, where $N=20$ is the number of spatial oscillations in the extended pattern. The existence intervals of these patterns depend on $n$, with patterns with consecutive values of $n$ overlapping while shifting to larger values of $a$ as $n$ increases as in Fig.~\ref{FigBif}.

\begin{figure}[t]
\centering
\includegraphics[width=1.0\columnwidth]{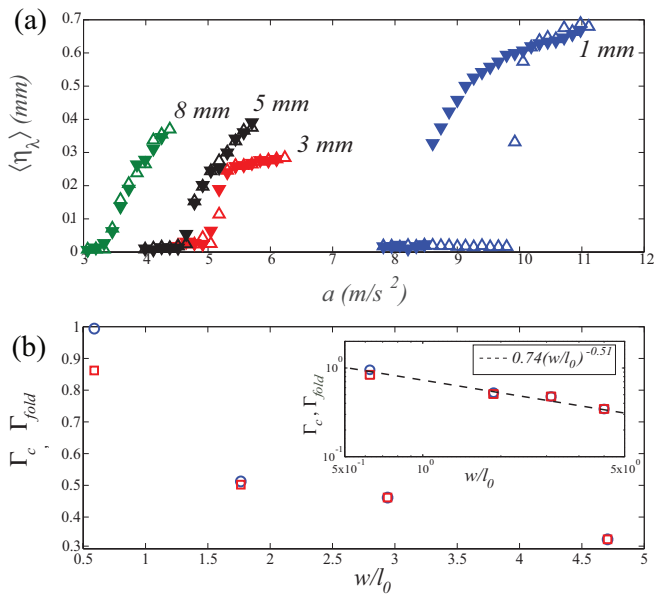} 
\caption{(color online) (a) Experimental bifurcation diagrams for extended states in a plexiglass cell for increasing ($\blacktriangle$) and decreasing ($\triangledown$) acceleration $a$ when $w$ = 1, 3, 5 and 8 mm. (b) Normalized acceleration at onset  $\Gamma_c=a_c/g$ ($\circ$) and at the fold $\Gamma_{fold}=a_{fold}/g$  ($\square$) as functions of the width $w$ normalized by the capillary length $l_0=$1.7 mm. Inset: Log-log plot of $\Gamma_c=a_c/g$ ($\circ$) and $\Gamma_{fold}=a_{fold}/g$ as a function of $w/l_0$.}
\label{FigWidth}
\end{figure}

The physical reason for the appearance of bistability and localized patterns outside this region is the contact angle hysteresis at the surface of the container. To probe this idea we have performed runs in plexiglass cells with the same $l=200$ mm, $h=40$ mm, $h_0=23$ mm and different widths $w$=1, 3, 5 and 8 mm. The bifurcation diagrams are shown in Fig.~\ref{FigWidth}(a): as $w$ increases the threshold $a_c$ for the parametric instability and the lowest acceleration $a_{fold}$ at which extended waves are still present both decrease (Fig.~\ref{FigWidth}(b)), and for $w\gtrsim 3$ mm the hysteresis and the localized waves both disappear. The up-down asymmetry of the extended pattern also decreases.

Slanted snaking of stationary localized states has been reported before~\cite{Firth2007b,Dawes2008,Beaume2013} and likewise attributed to the presence of a conserved quantity~\cite{Dawes2010,Knobloch2016}, following earlier work of Matthews and Cox~\cite{CM00,CM01}. The conserved quantity responsible for slanted snaking in the present system is the volume of fluid in the container.  The volume per unit length (i.e., the cross-sectional area) for a given meniscus height depends on the contact angle at the cross-stream boundaries and the associated angle hysteresis allows for the redistribution of fluid volume along the length of the cell as the contact angle changes.  The effect of this redistribution becomes more pronounced with a narrower cell because the volume of fluid associated with this motion constitutes a larger change in height when averaged across the width of the cell. In particular, the experiment reveals a slight decrease in the fluid level relative to the rest state outside of a localized Faraday wave ($\sim$100 $\mu$m for localized structures with 5-8 bumps), implying that the oscillatory motion of the interface within the wave acts to draw the conserved fluid away from the background flat state and into the wave.

We write the position of the liquid surface in the form
\begin{align}
h(x,t) =& h_0 + \left(A(X,T)e^{ik_c x}+\bar{A}(X,T)e^{-ik_c x}\right)e^{i\omega t/2}\nonumber\\
& + B(X,T) + c.c., \label{eq:height} 
\end{align}
where $A(X,T)$ is the (complex) amplitude of a surface wave oscillating with frequency $\omega/2$ at wave vector $k_c=k_c(\omega/2)$ and $B(X,T)$ is the (real) amplitude of a conserved mode that models the effects of contact angle hysteresis at the transverse walls on the height of the fluid surface. The resulting expression is to be interpreted as describing the width-averaged data. Using symmetry arguments or multiscale analysis~\cite{CM00,CM01,Knobloch2016} one finds that the simplest set of equations for the amplitudes $A$ and $B$ takes the (scaled) form
\begin{align}
A_T =& (\mu+i\nu)A+\gamma \bar{A}+(1+i\alpha)A_{XX} \nonumber \\ 
&-(1+i\beta)|A|^{2}A +(\delta+i\sigma) BA, \label{eq:fcgle}\\
B_T =& B_{XX} - (|A|^{2})_{XX}.\label{eq:cmode}
\end{align}

\begin{figure*}[t]
\centering
\includegraphics[width=1.0\textwidth]{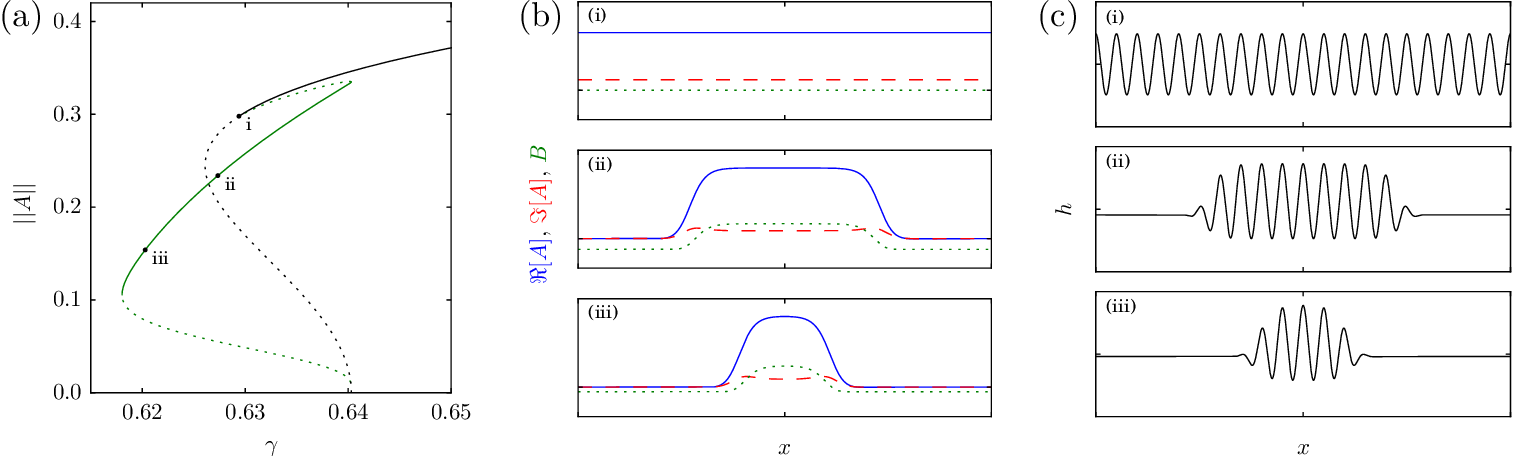} 
\caption{(color online) (a) Numerical bifurcation diagram showing time-independent solutions of Eqs.~(\ref{eq:fcgle})--(\ref{eq:cmode}). The homogeneous state (black) corresponds to a domain-filling Faraday wave while the localized states (green) lie on a slanted branch. Stable (unstable) portions of the solution branches are drawn with solid (dotted) lines. (b) Profiles of the real (blue solid line) and imaginary (red dashed line) parts of the amplitude $A$ and of the conserved mode $B$ (green dotted line) for both localized and homogeneous states. (c) Profiles of the elevation $h(x,0)$ of the fluid surface relative to $h_0$ as predicted by Eq.~(\ref{eq:height}); both odd and even states lie in the same branch. Parameters: $\mu=-0.5$, $\nu=0.4$, $\alpha=1$, $\beta=2$, $\delta=0.25$, $\sigma=0$, $\lambda_c=1$.}
\label{FigModel}
\end{figure*}

Equation~(\ref{eq:fcgle}) describes the long-time $T$ and large-scale $X$ evolution of the envelope $A(X,T)$ of a Faraday wave in a system that is parametrically driven near a $2:1$ temporal resonance \cite{Burke2008classification}. Here $\mu<0$ represents the effects of viscous damping, $\nu$ corresponds to detuning between $\omega/2$ and the natural frequency of oscillation at wavenumber $k_c$, and $\alpha$, $\beta$ and $\gamma$ measure dispersion, amplitude dependence of the frequency and the magnitude of the time-periodic forcing, respectively. The large scale mode $B(X,T)$ locally alters both the decay rate of free oscillations and the detuning, while the presence of the Faraday wave results in a redistribution of the conserved $B$ toward (or away from) regions with a large amplitude pattern. 

In the stationary state the amplitude $A(X)$ satisfies the nonlocal equation
\begin{align}
(1+i\alpha) A_{XX}+ (\mu + i\nu)A - (1+i\beta)|A|^2 A  & \nonumber \\
 +\gamma \bar{A} + (\delta+i\sigma)\left(|A|^2-\langle|A|^2\rangle\right) A & = 0, \label{eq:nonlocalGL}
\end{align}
where $\langle|A|^2\rangle\equiv l^{-1}\int_{-l/2}^{l/2}|A|^2 dX $. 
The trivial $A=0$ state corresponds to a flat surface with no pattern while the constant amplitude state $A=A_0\ne0$ corresponds to a domain-filling Faraday wave. Localized Faraday waves are represented by time-independent states that consist of a section of the $A=A_0$ state embedded in an $A=0$ background. In the absence of coupling to the large scale mode ($\delta=\sigma=0$) such states exist at a single value of $\gamma$. The coupling is responsible for effective dissipation that depends on the spatial extent of the localized state thereby generating a slant in the branch of localized solutions as a function of the forcing strength $\gamma$: broader solutions feel larger dissipation and require stronger forcing for their existence. Figure~\ref{FigModel} shows stationary localized and domain-filling patterns predicted by the amplitude equations (\ref{eq:fcgle})-(\ref{eq:cmode}) and confirms that the model equations not only capture qualitatively the predicted slanting of the branch of localized states but also the associated increase in water level as measured by the $B$ field. 

\begin{figure}[t]
\centering
\includegraphics[width=1.0\columnwidth]{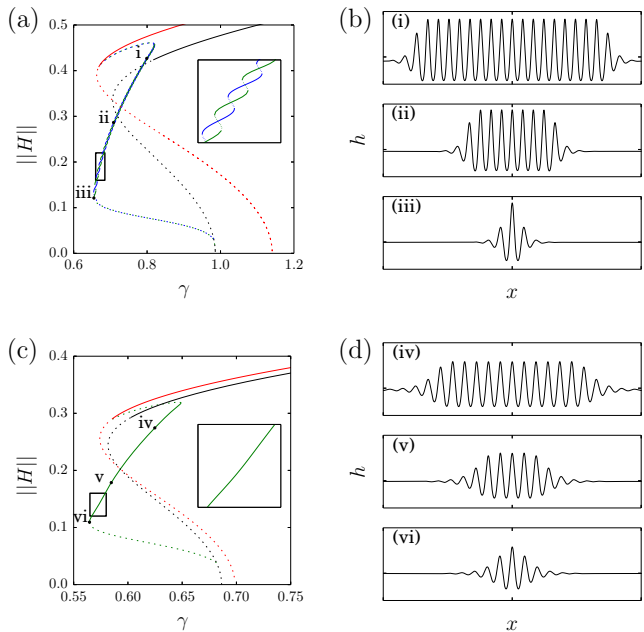} 
\caption{(color online) (a) Numerical bifurcation diagram showing time-independent solutions of Eqs.~(\ref{eq:fcgle_she}) and (\ref{eq:cmode}). The black curve represents the primary branch of domain-filling Faraday waves. Localized states (odd: green; even: blue) bifurcate simultaneously at small amplitude, extend beyond the fold and exhibit slanted snaking before terminating simultaneously on a different branch of domain-filling Faraday waves (red), cf.~\cite{Lojacono}. Solid (broken) lines indicate stable (unstable) solutions. (b) Profiles of the elevation $h(x,t)=h_0+B+H\exp(i\omega t/2) +c.c.$ for odd states at $t=0$ at labeled locations in (a). (c) For other parameter values smooth snaking is present (only odd localized states are shown). (d) Profiles of the fluid elevation $h(x,0)$ as described in (b). Parameters: $\mu=-0.5$, $\nu=0$, $\alpha=1$, $\beta=-4$, $\delta=1$, $\sigma=0$, $k_c=1$ and (a,b) $\zeta=1$, (c,d) $\zeta=0.5$. }
\label{FigModel2}
\end{figure}

Equations (\ref{eq:fcgle})-(\ref{eq:cmode}) describe small amplitude states near onset of the Faraday instability on the assumption that the envelope of the pattern varies on a length scale that is much longer than the characteristic wavelength $2\pi/k_c$. The resulting separation of the spatial dependence of the solution into dependence on slow and fast spatial scales eliminates pinning of the fronts on either side of the localized structure to the spatial oscillations within \cite{Knobloch2015,KaoK,Clerc}, thereby collapsing the branches of odd and even states into a single inclined nonsnaking branch. True snaking, such as that observed in our experiment, is only observed when terms beyond all orders are included in the amplitude equation \cite{Kozyreff} or, more usefully, in models in which the finite characteristic wave number $k_c$ is built into the equation for the fluctuating part of the wave profile, $\delta h(x,t)=H(x,t)\exp (i\omega t/2) + c.c.$, such as the parametrically forced complex Swift-Hohenberg equation, cf.~\cite{Malomed,GelensK},
\begin{eqnarray}
H_t =& (\mu+i\nu) H +i\zeta H_{xx}-(1+i\alpha)(\partial^2_x+k_c^2)^2H+\gamma \bar{H} \nonumber \\ 
&-(1+i\beta)|H|^{2}H +(\delta+i\sigma) BH, \label{eq:fcgle_she}
\end{eqnarray}
coupled to a conserved quantity as in Eq.~(\ref{eq:cmode}). As shown in Fig.~\ref{FigModel2}(a) odd and even localized solutions of this model do exhibit the type of intertwined slanted snaking observed in the experiment (Fig.~\ref{FigBif}), even though for smaller values of the parameter $\zeta$ the snaking is replaced by smooth snaking of the type observed in \cite{Dawes2010,Lojacono}. Larger values of $\zeta$ increase the slanted snaking range and move it further outside the hysteresis loop. However, in all cases the localized Faraday waves originate in a secondary bifurcation from a spatially extended Faraday wave state (Figs.~\ref{FigModel2}(b,d)). Note that detuning ($\nu\ne0$) is not necessary for this behavior. Our model suggests that the localized states shown in Fig.~\ref{FigBif} lie on connected intertwined branches with stable odd and even states separated by unstable states as in Fig.~\ref{FigModel2}(a).

Localized Faraday oscillations in Newtonian fluids confined in a Hele-Shaw cell have been observed before \cite{Rajchenbach2011} but their origin and possible snaking behavior was not studied. The present experiment reveals a large multiplicity of spatially localized Faraday waves organized in a slanted snaking bifurcation diagram, an observation that can be qualitatively understood using both amplitude equations (Eq.~(\ref{eq:fcgle})) and a model equation constructed in the spirit of the Swift-Hohenberg model (Eq.~(\ref{eq:fcgle_she})), provided that volume conservation is taken into account in both cases (Eq.~(\ref{eq:cmode})).

\acknowledgements{The authors acknowledge financial support from CONICYT grant CONICYT-USA PII20150011, and from FONDECYT grants 1130354 and 1150507 as well as the Berkeley-Chile Fund. The authors thank L. Gordillo and G. Camel for discussions.}


\begin{thebibliography}{29}

\bibitem{Descalzi2011} O. Descalzi, M. Clerc, S. Residori, and G. Assanto (eds), {\it Localized States in Physics: Solitons and Patterns} (Springer, New York, 2011).

\bibitem{Purwins2010}  H. G. Purwins, H. U. B\"odeker, and S. Amiranashvili, Adv. Phys. {\bf 59}, 485 (2010).

\bibitem{CrossHohenberg1993} M. C. Cross and P. C. Hohenberg, Rev. Mod. Phys. {\bf 65}, 851 (1993).
  
\bibitem{Knobloch2015} E. Knobloch, Annu. Rev. Cond. Matter Phys. {\bf 6}, 325 (2015).

\bibitem{Coullet2000} P. Coullet, C. Riera, and C. Tresser, Phys. Rev. Lett. {\bf 84}, 3069 (2000).

\bibitem{Beck} M. Beck, J. Knobloch, D. J. B. Lloyd, B. Sandstede and T. Wagenknecht, SIAM J. Math. Anal. {\bf 41}, 936 (2009).
  
\bibitem{Burke2006} J. Burke, and E. Knobloch, Phys. Rev. E  {\bf 73}, 056211 (2006).

\bibitem{Burke2007} J. Burke, and E. Knobloch, Chaos {\bf 17}, 037102 (2007).

  
  

\bibitem{Firth2007} W. J. Firth, L. Columbo and T. Maggipinto, Chaos {\bf17}, 037115 (2007).

\bibitem{Barbay2008} S. Barbay, X. Hachair, T. Elsass, I. Sagnes, and R. Kuszelewicz, Phys. Rev. Lett. {\bf 101}, 253902 (2008).

\bibitem{Abshagen2010} J. Abshagen, M. Heise, G. Pfister, and T. Mullin, Phys. Fluids {\bf 22}, 021702 (2010).

\bibitem{Haudin2011} F. Haudin, R. G. Rojas, U. Bortolozzo, S. Residori, and M. G. Clerc, Phys. Rev. Lett. {\bf 107}, 264101 (2011).

\bibitem{Thiele2013} U. Thiele, A. J. Archer, M. J. Robbins, H. Gomez, and E. Knobloch, Phys. Rev. E {\bf 87}, 042915 (2013).

\bibitem{Lloyd2015} D. J. B. Lloyd, C. Gollwitzer, I. Rehberg, and R. Richter, J. Fluid Mech. {\bf 783}, 283 (2015).

\bibitem{Firth2007b} W. J. Firth, L. Columbo, and A. J. Scroggie, Phys. Rev. Lett. {\bf 99}, 104503 (2007).
  
\bibitem{Dawes2008} J. H. P. Dawes, SIAM J. Appl. Dyn. Syst. {\bf 7}, 186 (2008).

\bibitem{Assemat} P. Assemat, A. Bergeon and E. Knobloch, Fluid Dyn. Res. {\bf 40}, 852 (2008).
  
\bibitem{Clerc} M.~G. Clerc, C. Fern\'andez-Oto, and S. Coulibaly, Phys. Rev. E  {\bf 87}, 012901 (2013).
  
\bibitem{Alnahdi} A.~S. Alnahdi, J.~Niesen, and A.~M. Rucklidge, SIAM J. Appl. Dyn. Sys. {\bf 13}, 1311 (2014).

\bibitem{Faraday1831} M. Faraday, Phil. Trans. R. Soc. London {\bf 121}, 299 (1831).
  
\bibitem{DelRioJColloid1997} O. del Rio and A. W. Neumann, J. Colloid Int. Sci. {\bf 196}, 136 (1997).

\bibitem{DeGennesReview} P. G. de Gennes, Rev. Mod. Phys. {\bf 57}, 827 (1985).

\bibitem{MaciasPRE2013} J. E. Mac\'ias, M. G. Clerc, C. Falc\'on, and M. A. Garc\'ia-\~Nustes, Phys. Rev. E {\bf 88}, 020201(R) (2013).



\bibitem{Beaume2013} C. Beaume, A. Bergeon, and E. Knobloch, Phys. Fluids {\bf 25}, 024105 (2013).

\bibitem{Dawes2010} J. H. P. Dawes and S. Lilley, SIAM J. Appl. Dyn. Syst. {\bf 9}, 238 (2010).

\bibitem{Knobloch2016} E. Knobloch, IMA J. Appl. Math. {\bf 81}, 457 (2016).

\bibitem{CM00} P. C. Matthews and S. M. Cox, Nonlinearity {\bf 13}, 1293 (2000).
  
\bibitem{CM01} S. M. Cox and P. C. Matthews, Physica D {\bf 149}, 210 (2001).

\bibitem{Burke2008classification} J. Burke, A. Yochelis and E. Knobloch, SIAM J. Appl. Dyn. Sys. {\bf 7}, 651 (2008). 
  
\bibitem{KaoK} H.-C. Kao and E. Knobloch, Phys. Rev. E  {\bf 85}, 026211 (2012).
  
\bibitem{Kozyreff} G. Kozyreff and S. J. Chapman, Phys. Rev. Lett. {\bf 97}, 044502 (2006).

\bibitem{Malomed} B. A. Malomed, Z. Phys. B {\bf 55}, 241 (1984).

\bibitem{GelensK} L. Gelens and E. Knobloch, Phys. Rev. E  {\bf 84}, 056203 (2011).
  
\bibitem{Lojacono} D. Lo Jacono, A. Bergeon and E. Knobloch, J. Fluid Mech. {\bf 687}, 595 (2011).

\bibitem{Rajchenbach2011} J. Rajchenbach, A. Leroux, and D. Clamond, Phys. Rev. Lett. {\bf 107}, 024502 (2011).

\end{thebibliography}
\end{document}